\documentclass[aps,prl,amsmath,twocolumn]{revtex4}
\usepackage{bm}
\usepackage{amssymb}
\usepackage{graphicx}
\newcommand{\dd}{\mathrm{d}}
\newcommand{\vc}[1]{{\bm{#1}}}

\begin{document}
\title{Spectral energy dynamics in magnetohydrodynamic turbulence}
\author{Wolf-Christian M\"uller}
\affiliation{Max-Planck-Institut f\"ur Plasmaphysik, 85748 Garching, Germany }
\author{Roland Grappin}
\affiliation{Observatoire de Paris-Meudon, 92195 Meudon, France}
\begin{abstract}
{Spectral direct numerical simulations of incompressible MHD turbulence at
a resolution of up to $1024^3$ collocation points are presented for a
statistically isotropic system as well as for a setup with an imposed strong
mean magnetic field.  The spectra of residual energy,
$E_k^\mathrm{R}=|E_k^\mathrm{M}-E_k^\mathrm{K}|$, and total energy,
$E_k=E^\mathrm{K}_k+E^\mathrm{M}_k$, are observed to scale self-similarly in the
inertial range as $E_k^\mathrm{R}\sim k^{-7/3}$, $E_k\sim k^{-5/3}$ (isotropic
case) and $E^\mathrm{R}_{k_\perp}\sim k_\perp^{-2}$, $E_{k_\perp}\sim
k_\perp^{-3/2}$ (anisotropic case, perpendicular to the mean field direction).
A model of dynamic equilibrium between kinetic and magnetic energy, based on the
corresponding evolution equations of the eddy-damped quasi-normal Markovian
(EDQNM) closure approximation, explains the findings. The assumed interplay of
turbulent dynamo and Alfv\'en effect yields $E_k^\mathrm{R}\sim k E^2_k$ which
is confirmed by the simulations.}  
\end{abstract}
\maketitle 
The nonlinear behavior of turbulent
plasmas gives rise to a variety of dynamical effects such as self-organization
of magnetic confinement configurations in laboratory experiments
\cite{ortolani_schnack:rfpbook}, generation of stellar magnetic fields
\cite{zeldovich:book} or structure formation in the interstellar medium
\cite{biskamp:book3}.  The understanding of these phenomena is incomplete as the
same is true for many inherent properties of the underlying turbulence.

Large-scale low-frequency plasma turbulence is treated in the
magnetohydrodynamic (MHD) approximation describing the medium as a viscous and
electrically resistive magnetofluid neglecting additional
kinetic effects.  Incompressiblity of the flow is assumed for the sake of
simplicity.  In this setting the nature of the 
turbulent energy cascade is a central and still debated issue 
with different phenomenologies
being proposed \cite{kolmogorov:k41a,iroshnikov:ikmodel,kraichnan:ikmodel,
sridhar_goldreich:gs1,goldreich_sridhar:gs3} (cf. \cite{mueller_biskamp:phenoreview} for a review). 
The associated spectral dynamics of kinetic and 
magnetic energy, in spite of its comparable importance, has received less attention 
(as an exception see \cite{grappin_pouquet_leorat:correl}).
    
This Letter reports a spectral relation between residual and
total energy, $E_k^\mathrm{R}=|E_k^\mathrm{M}-E_k^\mathrm{K}|$ and 
$E_k=E^\mathrm{K}_k+E^\mathrm{M}_k$ respectively,
as well as the influence of an imposed mean magnetic field on the spectra.
The proposed physical picture, which is
confirmed by accompanying direct numerical simulations, embraces two-dimensional 
MHD turbulence, globally isotropic three-dimensional systems as well as turbulence 
permeated by a strong mean magnetic field.

In the following reference is made to two high-resolution pseudospectral
direct numerical simulations of incompressible MHD turbulence which we regard as
paradigms for isotropic (I) and anisotropic (II) MHD turbulence.
The dimensionless MHD equations 
\begin{eqnarray}
\partial_t\vc{\omega}&=&\nabla\times\left[\vc{v}\times\vc{\omega}-\vc{b}\times
\left(\nabla\times\vc{b}\right) \right] + \mu\Delta\vc{\omega} \label{firstmhd}\\ 
\partial_t\vc{b}&=&\nabla\times\left(\vc{v}\times\vc{b}\right)+\eta\Delta\vc{b}\\
\nabla\cdot\vc{v}&=&\nabla\cdot\vc{b}=0\,. \label{lastmhd}
\end{eqnarray}
are solved in a $2\pi$-periodic cube with
spherical mode truncation to reduce numerical aliasing errors \cite{vincent_meneguzzi:simul}.
The equations include the flow vorticity, $\vc{\omega}=\nabla\times\vc{v}$, 
the magnetic field expressed in Alfv\'en speed
units, $\vc{b}$, as well as dimensionless viscosity, $\mu$, and resistivity,
$\eta$. {
In simulation II forcing is applied 
by freezing the largest spatial scales of velocity and magnetic field.}

Simulation I evolves globally isotropic freely decaying turbulence
represented by $1024^3$ Fourier modes.  The initial fields are smooth
with random phases and fluctuation amplitudes following
$\exp(-k^2/(2k_0^2))$ with $k_0=4$. Total kinetic and magnetic energy
are initially equal with $E^{\textrm{K}}=E^{\textrm{M}}=0.5$.  The
ratio $E^\textrm{K}/E^\textrm{M}$ decreases in time taking on values
of $0.28-0.23$ in the period considered
(cf. \cite{biskamp_mueller:3dmhddecay}). The ratio of kinetic and
magnetic energy dissipation rate, $\varepsilon^K/\varepsilon^M$, 
with $\mu=\eta=1\times 10^{-4}$ also
decreases during turbulence decay from $0.7$ to about $0.6$, the
difference in dissipation rates reflecting the imbalance of the
related energies.  The Reynolds number
\textsf{Re}$=(E^\mathrm{K})E/
(\mu\varepsilon^\mathrm{total})$ at $t=6$ is about $2700$ and
slightly diminishes during the run. 
Magnetic,
$H^{\mathrm{M}}=\frac{1}{2}\int_\mathrm{V}\dd\mathrm{V}\vc{a}\cdot\vc{b}$,
$\vc{b}=\nabla\times\vc{a}$, and cross helicity,
$H^\mathrm{C}=\frac{1}{2}\int_\mathrm{V}\dd\mathrm{V}\vc{v}\cdot\vc{b}$
, are negligible with $H^\mathrm{C}$ showing a dynamically unimportant
increase from 0.03 to 0.07 during the simulation.  The run covers 9
eddy turnover times defined as the time required to reach the maximum
of dissipation from $t=0$. The large-scale rms magnetic field
decays from initially 0.7 to 0.3.

Case II is a $1024^2\times 256$ forced turbulence simulation with an
imposed constant mean magnetic field of strength $b_0=5$ in units of
the large-scale rms magnetic field $ b_\mathrm{rms}\simeq v_\mathrm{rms}\simeq 1$. The forcing, which
keeps the ratio of fluctuations to mean field approximately constant,
is implemented by freezing modes with $k \le k_f=2$.  The simulation
with $\mu=\eta=9\times 10^{-5}$ has been brought into
quasi-equilibrium over 20 eddy-turnover times at a resolution of
$512^2\times 256$ and spans about 5 eddy turnover times of
quasi-stationary turbulence with $1024^2\times 256$ Fourier modes
and \textsf{Re}$\approx$$2300$ (based on field perpendicular
fluctuations).  Kinetic and magnetic energy as well as the ratio
$E^\mathrm{K}/E^\mathrm{M}$ are approximately unity with a slight
excess of $E^\mathrm{M}$. Perpendicular to the imposed field,
large-scale magnetic fluctuations with $b_\mathrm{rms}\simeq0.4$ are observed.  Correspondingly,
$\varepsilon^K/\varepsilon^M\simeq 0.95$ during the simulation. The
system has relaxed to $H^\mathrm{C}\simeq 0.15$ with a
fluctuation level of about 30\% and $H^\mathrm{M}\simeq
0.2H^\textrm{M}_\textrm{Max}$ with $H^\textrm{M}_\textrm{Max}\sim
E^\mathrm{M}/k_f$.

\begin{figure}
\centerline{\includegraphics[width=0.5\textwidth]{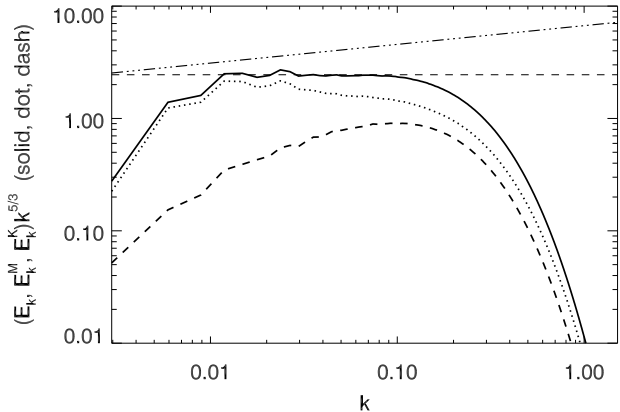}} 
\caption{Total (solid), kinetic (dashed), and magnetic (dotted) 
energy spectra in $1024^3$ case I simulation (normalized, time-averaged and compensated).
Dash-dotted line: $k^{-3/2}$ scaling.}
\label{f1} \end{figure}

Fourier-space-angle integrated spectra of total, magnetic, and kinetic 
energy for case I are shown in Fig. \ref{f1}. 
To neutralize secular changes as a consequence of turbulence decay,
amplitude normalization is used assuming a Kolmogorov total energy spectrum,
$E_k\rightarrow E_k/(\varepsilon \mu^5)$, $\varepsilon=-\partial_t E$,
with wavenumbers given in inverse multiples of the associated
dissipation length, $\ell_\mathrm{D}\sim
(\mu^3/\varepsilon)^{1/4}$. The quasi-stationary normalized spectra
are time averaged over the period of self-similar decay,
$t=6-8.9$.  As in previous numerical work
\cite{mueller_biskamp:3dmhdscale,haugen_brandenburg:3dmhdsimfull} and
also observed in solar wind measurements
\cite{leamon_etal:solwindspec,goldstein_roberts:solwind}, Kolmogorov
scaling applies for the total energy in the well-developed inertial
range, $0.01\alt k\alt 0.1$.  However, here the remarkable growth of
excess magnetic energy with decreasing wavenumber is of interest.
Qualitatively similar behavior is observed with large scale forcing
exerted on the system. We note that no pile-up of energy is
seen at the dissipative fall-off contrary to other high-resolution
simulations
\cite{haugen_brandenburg:3dmhdsimfull,kaneda_etal:hydrosim4096}. Apart
from different numerical techniques and physical models 
this difference might be due to the limited simulation
period at highest resolution namely 5
\cite{haugen_brandenburg:3dmhdsimfull} and 4.3
\cite{kaneda_etal:hydrosim4096} 
large-eddy-turnover times. {Depending on initial conditions 
the energy spectrum at
$1024^3$-resolution is still transient at that time.

In case II, pictured in Fig. \ref{f2}, 
\begin{figure}
\centerline{\includegraphics[width=0.5\textwidth]{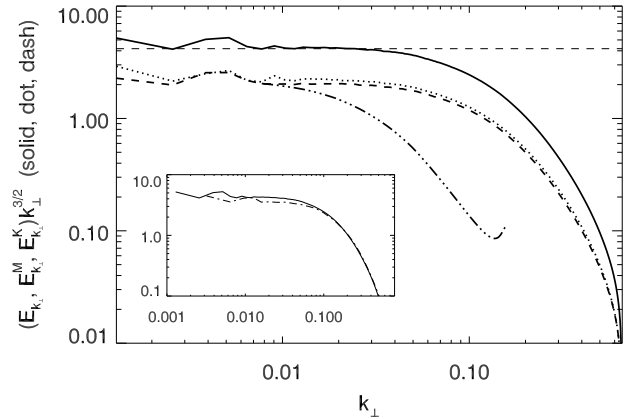}} 
\caption{Field-perpendicular total (solid), kinetic (dashed), and magnetic (dotted) 
energy spectra (normalized, time-averaged, and compensated) 
in $1024^2\times256$ case II simulation with $b_0=5$.
Dash-dotted curve: high-k part of  
field-parallel total energy spectrum. 
Inset: perpendicular total energy spectrum for 
resolutions of $512^2$ (dash-dotted) to $1024^2$ (solid).  
}
\label{f2} \end{figure}
strong anisotropy is generated due to turbulence depletion along the mean magnetic field,
$\vc{b}_0$, (cf. also \cite{mueller_biskamp_grappin:anisomhd,kinney_mcwilliams:anisomhd,oughton_matthaeus:anisormhd,
grappin:aniso,shebalin_matthaeus:aniso}).
This is visible when comparing the normalized and time-averaged field-perpendicular
one-dimensional spectrum, $E_{k_\perp}=\int\int\dd k_1 \dd k_2
E(k_\perp,k_1,k_2)$ (solid line) with the field-parallel spectrum, defined
correspondingly and adumbrated by the dash-dotted line in Fig. \ref{f2}.
The fixed $k_\perp$-axis is chosen arbitrarily in the
$k_1$-$k_2$-plane perpendicular to $\vc{b}_0$
where fluctuations are nearly isotropic. For the strong $b_0$ chosen here, field-parallel
and -perpendicular energy spectra do not differ notably from the ones
found by considering the direction of the local magnetic field as done e.g. in
\cite{mueller_biskamp_grappin:anisomhd,
cho_lazarian_vishniac:anisomhd}.
The field-parallel dissipation length is larger than in field-perpendicular directions
because of the stiffness of magnetic field lines. 
The numerical resolution in the parallel direction can, therefore, be reduced.

While there is no discernible inertial range in the parallel spectrum, its
perpendicular counterpart exhibits an interval with Iroshnikov-Kraichnan scaling,
$E_{k_\perp}\sim k_\perp^{-3/2}$ (Note that due to identical energy scales in 
Figs. \ref{f1} and \ref{f2} the absolute difference between  Kolmogorov and Iroshnikov-Kraichnan 
scaling is the same as in Fig.\ref{f1}).  
This is in contradiction to the anisotropic cascade
phenomenology of Goldreich and Sridhar for strong turbulence predicting
$E_{k_\perp}\sim k_\perp^{-5/3}$ \cite{goldreich_sridhar:gs3} and with numerical studies 
claiming to support the GS picture \cite{cho_vishniac:anisomhd,
cho_lazarian_vishniac:anisomhd}. However, the strength of $\vc{b}_0$ in these
simulations is of the order of the turbulent fluctuations and consequently much
weaker than for the anisotropic system considered here.
We note that indication for field-perpendicular IK
scaling has been obtained in earlier simulations at lower resolution 
using a high-order hyperviscosity and
with a stronger mean component, $b_0/b\sim 3\times10^2$ \cite{maron_goldreich:anisomhd}.
The authors of the aforementioned paper, however, are unsure whether {they} observe 
a numerical artefact or physical behavior.
 
The strongly disparate spectral extent of field-parallel and -perpendicular fluctuations
suggests 
that Alfv\'en waves propagating along the mean field do not have a 
significant influence on the perpendicular energy spectrum 
(in the sense of Goldreich-Sridhar, cf. also \cite{kinney_mcwilliams:anisomhd}). 
Instead, the strong $\vc{b_0}$ constrains turbulence to quasi-two-dimensional 
field-perpendicular planes as is well known and has been shown for this particular
system \cite{mueller_biskamp_grappin:anisomhd}. 

Another intriguing feature of system II is that $E^\mathrm{K}_k\simeq E^\mathrm{M}_k$
with only slight dominance of $E^\mathrm{M}$
(cf. Fig. \ref{f2}) in contrast to the growing excess of spectral magnetic
energy with increasing spatial scale for case I.  
Since both states are
dynamically stable against externally imposed perturbations (as has been verified numerically), 
they presumably represent equilibria between two competing nonlinear processes:
field-line deformation by turbulent motions on the spectrally local  
time scale $\tau_\mathrm{NL}\sim \ell/v_\ell\sim\left(k^3E^\mathrm{K}_k \right)^{-1/2}$
leading to magnetic field amplification (turbulent small-scale dynamo) and
energy equipartition by
shear Alfv\'en waves with the characteristic time 
$\tau_\mathrm{A}\sim\ell/b_0\sim(kb_0)^{-1}$ (Alfv\'en effect). 
The conjecture can be verified via the EDQNM closure
approximation \cite{orszag:edqnm} which yields
evolution equations for kinetic and magnetic energy spectra 
\cite{pouqet_frisch:invcasc} by including a phenomenological 
eddy-damping term for third-order moments.
The spectral evolution equation for the {signed}\footnote{{The other definition of $E^\mathrm{R}_k$ 
involving the 
modulus operator avoids case differentiations since 
the applied dimensional analysis is unable to predict the sign of $E^\mathrm{R}$. 
However, the physical picture underlying Eqs. (6) and (7) implies 
$E_k^\mathrm{M} \ge E_k^\mathrm{K}$ as it expresses an equilibrium between magnetic energy amplification
and equipartition of $E^\mathrm{K}$ and $E^\mathrm{M}$.}} residual energy, 
$E^\mathrm{R}=E^\text{M}-E^\text{K}$, in the case of negligible cross helicity
reads \cite{grappin_etal:zpzmspectra}: 
\begin{eqnarray}
\left(\partial_t+\left(\mu+\eta\right) k^2\right)E^\mathrm{R}_k
&=\int_{\triangle}\text{d} p \text{d} q 
\Theta_{kpq} \left(T^\mathrm{R}_{\mathrm{res}}+T^\mathrm{R}_\mathrm{crs}
+T^\mathrm{R}_\mathrm{Dyn}\right) \label{edqnm_res}
\end{eqnarray}
with the spectral energy flux contributions
\begin{eqnarray*}
T^\mathrm{R}_\mathrm{res}&=m_{kpq}\frac{k^2}{p}E^\mathrm{R}_pE^\mathrm{R}_q+
                      r_{kpq}\frac{p^2}{q}E^\mathrm{R}_qE^\mathrm{R}_k\,,\\
T^\mathrm{R}_\mathrm{crs}&=-m_{kpq}pE_qE^\mathrm{R}_k-t_{kpq}pE^\mathrm{R}_qE_k
\,,\\ 
T^\mathrm{R}_\mathrm{Dyn}&=\frac{s_{kpq}}{k}\left(k^2E_pE_q-p^2E_qE_k\right).
\end{eqnarray*}
The geometric coefficients $m_{kpq}$, $r_{kpq}$, $s_{kpq}$, $t_{kpq}$,
a consequence of the solenoidality constraints (\ref{lastmhd}), are given  
in \cite{grappin_etal:zpzmspectra}. The \lq$\triangle$' restricts integration
to {wave vectors} $\vc{k}$, $\vc{p}$, $\vc{q}$ which form a triangle, i.e. to a domain
in the $p$-$q$ plane which is defined by $q=|\vc{p}+\vc{k}|$.
The time $\Theta_{kpq}$ is characteristic of the eddy damping of the nonlinear
energy flux involving wave numbers $k$, $p$, and $q$. 
It is defined phenomenologically but its particular form
does not play a role in the following arguments. 

Local triad interactions with $k\sim p\sim q$ are dominating the hydrodynamic 
turbulent energy cascade and lead to Kolmogorov scaling of the associated
spectrum (cf., for example, \cite{lesieur:book}). In contrast, the nonlinear interaction of Alfv\'en waves includes
non-local triads with, e.g., $k\ll p\sim q$. 
{In this case a simplified version of equation (\ref{edqnm_res}) can be 
derived: 
\begin{eqnarray}
\partial_t E^\mathrm{R}_k &= -\Gamma_kkE_k^\mathrm{R}\equiv
T^\mathrm{R}_\mathrm{Alf},
\label{edqnm_nonloc1}
\end{eqnarray}
with $\Gamma_k=\frac{4}{3}k\int_0^{ak}\text{d} q\Theta_{kpq}E_q^\mathrm{M}$
\cite{pouqet_frisch:invcasc} $\sim k E^M \Theta$.}

It is now assumed that the right hand side of (\ref{edqnm_res}) can be written as 
$T^\mathrm{R}_\mathrm{Alf}+T^\mathrm{R}_\mathrm{Dyn}$ \cite{grappin_pouquet_leorat:correl}.
This states that the residual energy is a result of a dynamic equilibrium 
between turbulent dynamo and Alfv\'en effect.   
For stationary conditions and in the inertial range,  
dimensional analysis of (\ref{edqnm_res}) and (\ref{edqnm_nonloc1}) yields
$k^3E_k^2 \sim k^2E^\mathrm{M}E_k^\mathrm{R}$
which can be re-written as
\begin{equation}
E_k^\mathrm{R}\sim k E_k^2\,.
\label{erekrelation}
\end{equation}
The relaxation time, $\Theta$, appears 
as a factor on both sides of the relation and, consequently, drops out. 
We note that with $\tau_\mathrm{A}\sim (kb_0)^{-1}$, where $b_0$ is the mean magnetic field
carried by the largest eddies, $b_0\sim (E^\mathrm{M})^{1/2}$,
and by re-defining $\tau_\mathrm{NL}\sim\ell/(v_\ell^2+b_\ell^2)^{1/2}\sim(k^3 E_k)^{-1/2}$
(for system II all involved quantities are based on field-perpendicular fluctuations)
relation (\ref{erekrelation}) can be obtained in the physically more instructive form 
\begin{equation}
E_k^\mathrm{R}\sim \left(\frac{\tau_\mathrm{A}}{\tau_\mathrm{NL}}\right)^2 E_k\,.
\label{erek2}
\end{equation}
The modification of $\tau_\mathrm{NL}$ is motivated by considering that gradients of the Alfv\'en
speed contribute to nonlinear transfer as much as velocity shear
(see, e.g., \cite{heyvaerts_priest:phasemixing}).

For the examined setups relation (\ref{erek2}) is consistent with the
underlying physical idea of dynamical equilibrium between Alfv\'en and
dynamo effect. {At small scales with $k\gg k_0$ (for system II:
$k_0\simeq k_f$), Alfv\'enic interaction always dominates the energy
exchange since $\tau_\mathrm{A}\ll\tau_\mathrm{NL}$ (e.g. at
$k=0.3l_\mathrm{D}^{-1}$ for system I: $\tau_\mathrm{A}\simeq 5\times
10^{-2}, \tau_\mathrm{NL}\simeq 0.2$, for system II:
$\tau_\mathrm{A}\simeq 1\times 10^{-2}, \tau_\mathrm{NL}\simeq 0.1)$
which results in approximate spectral equipartiton of kinetic and
magnetic energy.  At larger spatial scales the Alfv\'en effect becomes
less efficient in balancing the transformation of kinetic to magnetic
energy by the small-scale dynamo with
$\tau_\mathrm{A}\simeq\tau_\mathrm{NL}$ (e.g. at $k=0.01
l_\mathrm{D}^{-1}$for system I: $\tau_\mathrm{A}\simeq 0.9,
\tau_\mathrm{NL}\simeq 0.8$, at $k=3\times 10^{-3} l_\mathrm{D}^{-1}$
for system II: $\tau_\mathrm{A}\simeq 1.2, \tau_\mathrm{NL}\simeq
0.9)$ allowing larger deviations from equipartition.}  

An interesting consequence of (\ref{erekrelation}) is that the difference 
between possible spectral scaling exponents, which is typically small
and hard to measure reliably, is enlarged by a factor
of two in $E^\mathrm{R}_k$. Even with the limited Reynolds numbers
in today's simulations such a magnified difference is clearly observable (e.g. dash-dotted lines in Figs. 
\ref{f1} and \ref{f3}). 
For system I with Kolmogorov scaling, $E_k\sim k^{-5/3}$
(Fig. \ref{f1}), relation (\ref{erekrelation}) predicts
$E^\mathrm{R}_k\sim k^{-7/3}$ in agreement with the simulation
(Fig. \ref{f3}).  In the case of Iroshnikov-Kraichnan behavior,
$E_{k_\perp}\sim k_\perp^{-3/2}$ as realized in system II
(Fig. \ref{f2}), $E^\mathrm{R}_{k_\perp}\sim k_\perp^{-2}$ is
obtained.  This result is confirmed by the residual energy spectrum
shown in Fig. \ref{f4} (cf. also \cite{biskamp:residualspec} for
two-dimensional MHD simulations and
\cite{grappin_pouquet_leorat:correl} for spectral model calculations).

\begin{figure}
\centerline{\includegraphics[width=0.5\textwidth]{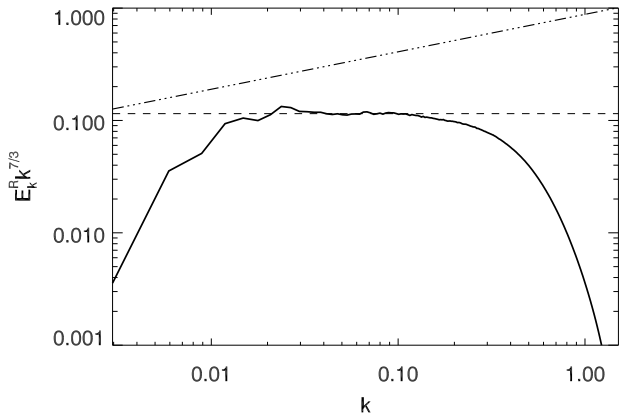}
} 
\caption{Compensated, space-angle-integrated residual energy spectrum,
$E^\mathrm{R}_{k}$, for same system as in Fig. \ref{f1}. Dash-dotted line:
$k^{-2}$-scaling.}
\label{f3}
\end{figure} 
\begin{figure} 
\centerline{ 
\includegraphics[width=0.5\textwidth]{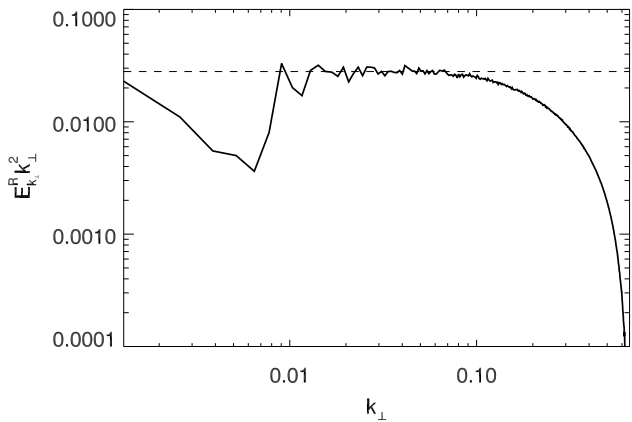} 
}
\caption{Compensated field-perpendicular residual energy spectrum
for the same system as in Fig. \ref{f2}.} \label{f4} 
\end{figure} 

In summary, based on the structure of the EDQNM closure equations for incompressible MHD 
a model of the nonlinear spectral interplay between kinetic and magnetic energy is formulated.
{Throughout the inertial range a quasi-equilibrium of turbulent small-scale dynamo and Alfv\'en effect
leads to the relation, $E^\mathrm{R}_k\sim kE_k^2$,}  linking total and residual energy spectra, in particular 
$E^\mathrm{R}_k\sim k^{-7/3}$ for $E_k\sim k^{-5/3}$ and
$E^\mathrm{R}_k\sim k^{-2}$ for $E_k\sim k^{-3/2}$.
Both predictions are
confirmed by high-resolution direct numerical simulations, 
limiting the possible validity of the Goldreich-Sridhar 
phenomenology to MHD turbulence with moderate mean magnetic fields.
\begin{acknowledgments}
The authors would like to thank Jacques L\'eorat and Dieter Biskamp for helpful discussions. WCM 
acknowledges financial support 
by the CNRS and CIAS, Paris Observatory.
\end{acknowledgments}
\bibliographystyle{/home/Wolf/.TeX/revtex4/apsrev}
\newcommand{\nop}[1]{}

\end{document}